\def\be{\begin{equation}}
\def\ee{\end{equation}}
\def\bea{\begin{eqnarray}}
\def\eea{\end{eqnarray}}

\documentclass[12pt]{article}
\usepackage{epsfig,latexsym}
\begin{document}
\thispagestyle{empty}
\vspace*{-0.5 cm}
\vspace*{-1.2in}
\vspace*{0.7 in}
\begin{center}
{\large \bf Proper time regulator and  Renormalization Group flow}
\\
\vspace*{1cm} 
{\bf M. Mazza}\\ \vspace*{.3cm} 
{\it Dipartimento di Fisica, Universit\`a di Catania}\\
{\it INFN, Sezione di Catania}\\
{\it Corso Italia 57, I-95129, Catania, Italy} \\
\vspace*{.6cm}
 and \\
\vspace*{.6cm}
{\bf D. Zappal\`a }\\ \vspace*{.3cm} 
{\it INFN, Sezione di Catania}\\
{\it Dipartimento di Fisica, Universit\`a di Catania}\\
{\it Corso Italia 57, I-95129, Catania, Italy} \\

\vspace*{1 cm}
{\bf ABSTRACT} \\
\end{center}
We consider some applications of the Renormalization Group
flow equations obtained by resorting to a specific class 
of proper time regulators.
Within this class a particular limit that corresponds to 
a sharpening of the effective width of the regulator is
investigated and a procedure to analytically implement 
this limit on the flow equations is shown.
We focus on the critical exponents determination for the 
$O(N)$ symmetric scalar theory in three dimensions. The large $N$ 
limit and some perturbative features in four dimensions
are also analysed. In all problems examined the results
are optimized when the mentioned limit of the proper time
regulator is taken.
\\
\vskip 0.5 cm
\noindent
Pacs 11.10.Hi , 11.10.Kk

\parskip 0.3cm
\vspace*{3cm}
\vfill\eject
\setcounter{page}{1}
\voffset -1in
\vskip2.0cm

\newcommand{\fa}{\phi^a}
\newcommand{\fb}{\phi^b}
\newcommand{\p}{\partial_{\mu}}
\newcommand{\dd}{\delta^{ab}}
\newcommand{\nn}{\nonumber}

\section{Introduction}

In the past years there has been growing interest for the Renormalization
Group (RG) techniques originally inspired to the Kadanoff-Wilson  blocking 
procedure \cite{kad},
as they represent promising means to investigate areas that are 
out of reach of the standard perturbative quantum field theory. 
The central idea is the description of the  RG flow in terms
of differential equations in the momentum space.
Some of the original works on this subject are \cite{weg,nicol,polch,hasen}.
More recently, a new formulation of the problem, 
usually called Exact Renormalization Group (ERG), has been 
developed \cite{wet1,mor1,mor2} (for recent reviews on the subject and on  various
applications see\cite{berg2,bagn}). The ERG equations are functional differential equations
which describe the momentum scale $k$ dependence of a particular functional, usually 
indicated as average action. This functional is in some sense similar to the effective
action, which is the Euclidean generator of the one particle irreducible correlation
functions, with the difference that in the former only fluctuations with momenta 
larger than $k$ are included. This separation between high and low frequencies
is realized by the introduction of a smooth cutoff regulator $R_k$.
A relevant property of this functional is that it interpolates between the
classical Euclidean action and the full effective action of the theory, when the 
scale $k$ is lowered from the ultraviolet cutoff, where the theory
is defined by the local classical action, down to zero where all the fluctuations
are integrated.

A slightly different approach to the RG flow has been formulated in \cite{ole,flo,sen1}
where an operator cutoff is introduced  by means of the Schwinger proper 
time regulator. The most remarkable feature of this regulator is that it is 
formulated in a gauge invariant way and it has been already employed 
to compute the one loop beta function of Yang-Mills theories\cite{sen2}.
Therefore it is a promising tool for any analysis involving gauge theories. 
Besides that, it has been used to derive flow equations for  scalars 
coupled to fermions\cite{sha1} and for $O(N)$ scalar theories\cite{sha2}.
The differential equation that describes the flow of the scale dependent 
action $\Gamma_k$ obtained via the proper time regulator is
\be
\label{eq:gamma}
k\;{\partial \Gamma_k \over \partial k }=
-{1\over 2} {\rm Tr} \; \int_0^\infty \;{ds\over s} \;
k {\partial f_k \over \partial k}\;
{\rm exp}\; \Big (-s{\delta^2 \Gamma_k \over 
\delta \phi\delta \phi} \Big )   
\ee
with the constraints on the dimensionless cutoff function $f_k(x)$ that $f_k(0)=1$
and $f_k(x\to\infty)=0$ and the initial condition that at some ultraviolet scale $\Lambda$
$\Gamma_k$ is equal to the classical action $S$. The trace sums over all the 
discrete and continuous indices of the field $\phi$.
A connection to the perturbative one loop effective action 
\be
\label{eq:g1l}
\Gamma^{1-loop} = -{1\over 2} {\rm Tr} \; \int_0^\infty \;{ds\over s}  \; 
{\rm exp}\; \Big ({-s{\delta^2 S \over 
\delta\phi\delta\phi}} \Big )
\ee
is obtained if one neglects the effects of the running by freezing $\Gamma_k$ in the r.h.s. 
of Eq. (\ref{eq:gamma}) at its initial condition $S$. In this case $\Gamma_k$, due to 
the condition $f_k(0)=1$, coincides at $k=0$ with $\Gamma^{1-loop}$. 
The running quantity  $\Gamma_k$ in Eq. (\ref{eq:gamma}) corresponds to an infinite
resummation of diagrams and therefore improves on the one loop approximation. 
In practice the regulator $k \partial_k f_k$ acts as a particular smooth cutoff 
or weight function in the integration of the modes, selecting only the modes 
within a small shell centered around the scale $k$. 

However there is no direct connection to the smooth cutoff $R_k$, 
introduced in the ERG formalism (see for instance \cite{berg2}). 
As a consequence, the proof that in the limit $k\to 0$ 
the ERG flow of the average action $\Gamma_k$ converges to the 
usual effective action cannot be simply translated 
to the flow determined by the proper time regulator 
and a definite indication that this flow actually 
converges to the full effective action is missing.  

It should not be forgotten, however, that for practical 
purposes it is impossible to deal with the flow of the full 
action and, typically, a semi-local derivative expansion of 
the action is introduced to reduce the problem of the evaluation 
of the flow to a treatable  set of coupled partial differential 
equations. The best approximation analysed so far is the next to 
leading order in the derivative expansion which corresponds to 
take the following ansatz for $\Gamma_k$
\be
\label{eq:formgam}
\Gamma_k\lbrack\phi\rbrack=\int d^Dx \Big (V(\phi)+{{Z(\phi)}\over  {2}}\partial_\mu \phi
\partial^\mu\phi  \Big )
\ee
neglecting all the higher derivative terms, and determine the corresponding 
differential flow equation for the potential $V$ and for the 
wave function renormalization $Z$ (obviously $V$ and $Z$ depend on the 
scale $k$ but this dependence is not explicitly 
indicated to simplify the notation).
Due to this approximation, some uncertainty is introduced  about the 
$k\to 0$ limit of the truncated $\Gamma_k$ and a scheme dependence, 
basically related to the specific regulator employed, 
appears in the determination of the various physical 
quantities. This unphysical feature, which would disappear if one 
could deal with the flow of the full action, cannot be easily quantified. 
According to this remark, it is worthwhile to compare the 
predictions obtained with the proper time regulator  
(although lacking  of a convergence proof) with the results
obtained from the ERG. In \cite{io} it was noticed that
the following choice of the function $f_k$
\be
\label{eq:effe}
f_k(s Z k^2)=e^{-s Z k^2}\sum^m_{i=0}{{(sZ k^2)^i}\over {i!}},
\ee
parametrized in terms of the integer $m$,
when the value of $m$ is sufficiently large
provides a determination of
the critical exponents $\eta,\nu,\omega$
for the one-field scalar theory in $D=3$ dimensions
that is certainly comparable to the other determinations obtained
in the ERG framework.

In this paper we shall follow this point of view and explore
the possibility of explicitly studying the limit in which $m$
is infinite. It will be shown that for a simple particular 
parametrization it is possible to take the limit $m\to \infty$
and get two coupled equations for the potential and the 
wave function renormalization  which do not contain the parameter 
$m$ anymore. As noticed in \cite{io} for growing $m$ the 
proper time regulator selects smaller and smaller momentum
shells, centered around the scale $k$, which provide a relevant 
contribution to the flow differential equations. In this sense
the limit $m\to \infty$ corresponds to a procedure of 
reducing the characteristic width of a smooth cutoff and 
taking the sharp cutoff limit.
Due to the singular nature of this limit, the operation 
is not unique and depends on the specific smooth cutoff chosen.
The central result of this paper is that the equations obtained
when $m\to \infty$ yield a determination of the critical exponents
which actually agrees very well with the extrapolation of the 
corresponding values obtained at finite $m$, thus confirming 
that these universal quantities for large $m$ do not trivially 
diverge or vanish but rather have a finite sensible limit.
Moreover these values are very well compatible with other determinations
stemming from different, well established techniques (see \cite{zinnj}
or \cite{guida}).
In Sect. 2 we consider the coupled flow equations of $V$ and $Z$
for a scalar theory obtained with the proper time regulator and 
determine their form in the limit $m\to \infty$.
As a simple application we show that the perturbative two loop
anomalous dimension of the scalar field is easily recovered.
In Sect. 3 we collect the numerical results on the 
critical exponents at the non-trivial fixed point in $D=3$ dimensions 
for finite values of the parameter $m$ and in the limit $m\to \infty$.  
In Sect. 4 the particular case of the large $N$ limit of the potential 
equation for a scalar $O(N)$ symmetric theory is considered in order to 
check that the proper time regulator reproduces the known exact results.
The conclusions are summarized in Sect. 5.

\section{One component scalar theory}

Starting from the evolution equation (\ref{eq:gamma}), 
with the ansatz (\ref{eq:formgam}) for the action of a single scalar 
field theory in $D$ dimensions one can extract the flow equations 
for $V$ and $Z$.
The procedure is outlined in \cite{io} and 
the result is
\be
\label{eq:uflow}
k{{\partial V}\over{\partial k}} =\alpha k^D \Biggl({{Z k^2}\over{Z k^2+V''}}
\Biggr)^{m+1-D/2}
\ee
\eject
\bea
\label{eq:zflow}
&&k{{\partial Z}\over{\partial k}} =\alpha k^D
\Biggl({{Z k^2}\over{Z k^2+V''}}\Biggr)^{m+1-D/2}\Biggl\lbrack{{(m+1-D/2)}\over{Z k^2+V''}}
\biggl(-Z''
\nonumber\\
&&+{{(4+18D-D^2) (Z')^2}\over{24Z}}\biggr)
+{{(10-D)(m+1-D/2)(m+2-D/2)}\over{6 (Z k^2+V'')^2}}Z'V'''
\nonumber\\
&&-{{(m+1-D/2)(m+2-D/2)(m+3-D/2)}\over{6(Z k^2+V'')^3}}Z(V''')^2\Biggr \rbrack
\eea
where each prime indicates a derivative w.r.t. the field $\phi$ and 
the constant $\alpha$ is expressed in terms of gamma functions $\Gamma(x)$
\be
\label{eq:alfa}
\alpha={{\Gamma(m+1-D/2)}\over{(4 \pi)^{D/2} \Gamma(m+1)}}
\ee
One can immediately realize that the quantity $V''/(Z k^2)$ 
plays a fundamental role when the large $m$ limit of 
Eqs.(\ref{eq:uflow},\ref{eq:zflow}) is considered. 
In fact if $V''/(Z k^2)>0$ the term raised to the $m$-th power 
in (\ref{eq:uflow},\ref{eq:zflow}) vanishes, and if $V''/(Z k^2)<0$ 
it diverges (we are not interested here in the broken phase sector 
and $V''/(Z k^2)<-1$ is never considered). This however does not imply 
that the universal properties related to Eqs.(\ref{eq:uflow},\ref{eq:zflow}) must 
have a singular (or trivial) behavior for large $m$. Actually in \cite{io} 
it is found that already for values of $m$ around $10$, the 
critical exponents are almost $m$-independent. 

An interesting information about $V''$ can be read
in Fig.1 where the particular case of the non-Gaussian
fixed point in $D=3$ for the potential only (and $Z=1$ fixed) is examined
(more details about this calculation are given in Sect. 3).
The triangles in the upper frame of the figure correspond to the 
values of  $- V''(\phi=0)$ at the fixed point,
properly expressed as a dimensionless variable (see Sect. 3),
plotted for some values of $1/m$ (note that in 
this particular case the curvature  at the origin is negative). 
We consider  $1/m$ instead of $m$ because we are essentially interested 
in the large $m$ limit.
The circles in the upper frame correspond to the product $- m V''(\phi=0)$.
In the lower frame the circles are the same of the upper part of the figure,
but plotted on a magnified scale. It is evident  from Fig. 1 that
$V''(\phi=0)$ at the fixed point behaves, at least in first 
approximation, like $1/m$ for large $m$. Of course this does not 
simply mean that $m V''(\phi=0)$ is a constant. Rather, the 
plot in the lower frame suggests a behavior like 
 $- m V''(\phi=0)=a-b/m$ with $a,b$ positive constants (up to the accuracy considered).
However, our aim here is just to find out the leading behavior 
of $V''(\phi=0)$ for large $m$, which, as noticed above, is $1/m$.
Supposing that $V''$ has the trend of  Fig.1, not only at $\phi=0$ 
but at any value of $\phi$, it would be helpful to redefine the variables 
in Eqs.(\ref{eq:uflow},\ref{eq:zflow}) in such a way that this dependence 
on $m$ could be compensated.

As a first step we can get rid of the constant $\alpha$
in Eqs. (\ref{eq:uflow},\ref{eq:zflow}) by simply redefining
$V\to \alpha V$, $\phi \to \sqrt{\alpha}\phi$ and leaving $Z$
unchanged. According to Eq.(\ref{eq:formgam}), this operation
amounts to rescale the action by the constant $\alpha$.
Then, following the indications of Fig.1, we can make a 
second transformation by replacing  $\phi \to \sqrt{m}\phi$.
This time the action is not uniformly modified. With the help of
the latter substitution we can explicitly take the $m\to\infty$ 
limit of Eqs. (\ref{eq:uflow},\ref{eq:zflow}) and get 
\be
\label{eq:ufminf}
k{{\partial V}\over{\partial k}} = k^D e^{-V''/(Z k^2)}
\ee
\bea
\label{eq:zfminf}
&&k{{\partial Z}\over{\partial k}} =
 k^D e^{-V''/(Z k^2)}\nonumber\\
&&\times\Biggl(
-{{Z''}\over{Z k^2}}+{{(4+18D-D^2) (Z')^2}\over{24Z^2k^2}}
+{{(10-D)Z'V'''}\over{6 (Z k^2)^2}}-
{{Z(V''')^2}\over{6(Z k^2)^3}}\Biggr )
\eea
We see that the terms raised to the $m$-th power in
(\ref{eq:uflow},\ref{eq:zflow}) do not turn into  singularities,
but instead into exponential terms. In addition, other terms of Eq. 
(\ref{eq:zflow}) do not appear in Eq.(\ref{eq:zfminf}) because
they are suppressed for   $m\to\infty$.
Eqs. (\ref{eq:ufminf},\ref{eq:zfminf}) can now be used to study  
the critical properties.

A simple check on Eqs. (\ref{eq:uflow},\ref{eq:zflow}) and 
(\ref{eq:ufminf},\ref{eq:zfminf}) can be performed 
by looking at the perturbative determination in $D=4$ of the anomalous 
dimension $\eta$ of the field $\phi$, defined as 
$\eta=-k\partial_k {\rm ln}(\overline Z)$ where $\overline Z$ 
is the residue of the mass pole of the two point Green function 
of the theory. The specific truncation of the derivative expansion 
of the action above considered is actually sufficient to derive the 
lowest order non-vanishing perturbative contribution to $\eta$ 
(see for instance \cite{io2l} where the same calculation, in a 
different framework, namely starting from the Wegner-Houghton 
flow equation, has already been performed). In fact if one starts with 
the classical Euclidean action (we require the invariance under the 
transformation $\phi\to -\phi$)
\be
\label{eq:scl}
S={1\over 2}(\partial \phi)^2+{{\lambda\phi^4}\over {4!}}
\ee
then the lowest order values, in the perturbative parameter $\lambda$, 
of $Z$ and $V$ in (\ref{eq:uflow},\ref{eq:zflow}), are respectively $1$
and $\lambda\phi^4/4!$. If these lowest order values are inserted 
in Eq. (\ref{eq:zflow}) and the function $Z$ is conveniently expanded
in even powers of the field 
\be
\label{eq:zexpa}
Z=z_0+{{z_2\phi^2}\over 2}+{{z_4\phi^4}\over {4!}}+...
\ee
then, by comparing the terms proportional to $\phi^2$ in (\ref{eq:zflow}),
one immediately sees that $z_2$ is $O(\lambda^2)$. More precisely, by 
integrating the differential equation from the ultraviolet cutoff $\Lambda$
to the scale $k<<\Lambda$ (but still $k$ much larger than any other infrared 
scale such as the field mass pole)
\be
\label{eq:zdue}
z_2=
{{(m+1)}\over{ 16\pi^2}}{{\lambda^2}\over{ 6 k^2}}+O(\lambda^3)+O(1/\Lambda^2)
\ee

Consequently, $z_0$, which at the leading order is $z_0=1$ 
(as it follows from 
(\ref{eq:scl}) and (\ref{eq:zexpa})), to the next to leading order gets a 
$O(\lambda^2)$ contribution from $z_2$ that enters the flow equation of $z_0$
through the term $Z''$ in Eq. (\ref{eq:zflow}). We get
\be
\label{eq:zzero}
k {{\partial z_0}\over{k}}=-{{(m+1)}\over{6 m}}\left ({{\lambda}\over{ 16 \pi^2}}
\right )^2+O(\lambda^3)
\ee
 
The two point Green function is obtained from the double functional derivative 
of (\ref{eq:formgam}) at $\phi=0$. It is easy to realize that to the next to leading
order in the derivative expansion we get $\Gamma^{(2)}(p^2)=z_0 p^2+ V''(\Phi=0)$
and therefore, to this order, $\overline Z=z_0$. It follows
\be
\label{eq:eta}
\eta={{(m+1)}\over{6 m}}\left ({{\lambda}\over{ 16 \pi^2}}
\right )^2 +O(\lambda^3)
\ee
The important result here is that the perturbative value of $\eta$ (see e.g. \cite{zinnj}) corresponds to
the large $m$ limit of Eq. (\ref{eq:eta}), which is in agreement 
with our conjecture that
the physically relevant results are obtained for
$m\to\infty$.

One remark is to be made here about other terms neglected in the derivative
expansion (\ref{eq:formgam}) that are $O(\lambda^2)$. In fact terms like 
$Y\phi^2\partial^n\phi\partial^n\phi$ with $n\geq 2$, which are neglected to order
$O(\partial^2)$, can be computed with the same method employed to determine
Eqs.(\ref{eq:uflow},\ref{eq:zflow})and it turns out that $Y\propto \lambda^2/k^{2n}$.
However the contribution of $Y$ to $\Gamma^{(2)}(p^2)$ is proportional
to $\lambda^2(p^2/k^2)^n$ and, as long as we are interested in a momentum
range where $k^2$ is much larger than a mass pole, then $k^2>>p^2$ and, consistently,
we can neglect the contributions of $Y$ and similar terms which are suppressed
by the factor $(p^2/k^2)^n$.

Finally we turn to Eqs. (\ref{eq:ufminf},\ref{eq:zfminf}) and perform the same 
perturbative analysis. In order to get the same normalization used to derive
Eq. (\ref{eq:eta}), one has to remember that, globally, the following substitutions
have been performed to go from (\ref{eq:uflow},\ref{eq:zflow}) 
to (\ref{eq:ufminf},\ref{eq:zfminf}): $V\to\alpha V$, $\phi\to\sqrt{\alpha m}\phi$
and no rescaling on $Z$. As a consequence, instead of starting with the classical 
action $S$ in (\ref{eq:scl}) we have to start with a modified action 
$\widehat S$, equal in form to  (\ref{eq:scl}), but with $\lambda$
replaced with  $\widehat \lambda=\alpha m^2\lambda$, which for $m\to\infty$
becomes (see Eq. (\ref{eq:alfa}))
\be
\label{eq:lamcap}
\widehat \lambda={\lambda\over{16\pi^2}}
\ee
Note that since $Z$ is unchanged the initial value in (\ref{eq:scl}), $Z=1$, i.e. 
$z_0=1$, is unchanged too. A non-vanishing initial value for $z_2$ would have
required a rescaling according to the definition (\ref{eq:zexpa}).
After this remark, it is straightforward to derive $z_2$ and $\eta$ 
(through $z_0$) from Eq. (\ref{eq:zfminf}):
\be
\label{eq:zdue2}
z_2={{\widehat\lambda^2}\over{ 6 k^2}}+O(\widehat\lambda^3)+O(1/\Lambda^2)
\ee
\be
\label{eq:eta2}
\eta={{\widehat\lambda^2}\over{6}}+O(\widehat\lambda^3)
\ee
As expected the large $m$ limit of Eq. (\ref{eq:eta}) coincides with 
Eq. (\ref{eq:eta2}) which is the perturbative value of $\eta$.

\section{Numerical results}

In this Section we focus on the problem of determining,
through a numerical analysis, the fixed point solutions 
of our  flow equations and the  critical exponents $\nu$ 
and $\omega$ related to the eigenvalues of the linearized 
version of the flow equation around the fixed point.
In particular we are interested in the three dimensional 
case for which a large number of determinations of these 
universal quantities, obtained employing many different
techniques, is available in the literature (see e.g. \cite{zinnj,guida,berg2}).
Therefore, for the rest of this Section we set $D=3$.

The numerical determination of the exponents for this 
particular kind of proper time regulated flow, was started in \cite{io}.
There, $\nu$, $\omega$ and the anomalous dimension $\eta$ 
of the one component scalar theory,
determined by the numerical resolution of Eqs.
(\ref{eq:uflow}) and (\ref{eq:zflow}) are displayed for increasing 
values of the parameter $m$ and the data show a monotonous 
dependence on $m$. In particular $\nu$ and $\omega$
are determined up to $m=9$ and $\eta$ up to  $m=40$. 
Since in the second case $m$ is sufficiently large, 
the asymptotic value (for $m\to\infty$) of $\eta$ is obtained 
through a numerical fit in \cite{io}
and it is only claimed that $\nu$ and $\omega$ 
converge to finite values.

In the following we refine this analysis and evaluate $\eta$, $\nu$ and $\omega$
to the leading order ($O(\partial^0)$) in the derivative expansion 
(by solving the potential part
of the flow keeping $Z=1$ and $\eta=0$ fixed), and to the next to leading 
order ($O(\partial^2)$)
(by solving the full coupled problem for the potential and the wave 
function renormalization). In most cases we push the parameter $m$ up 
to $m=120$, which clearly shows the numerical trend of
the exponents and then we compare the obtained  values with the
ones determined from the equations (\ref{eq:ufminf}) and (\ref{eq:zfminf})
which hold for  $m\to \infty$.
We also extend the analysis to the $O(N)$ symmetric  theory
limiting ourselves to the $O(\partial^0)$ approximation 
in the derivative expansion
and derive $\nu$ and $\omega$ for $N=2,3,4$,
for increasing values of $m$ and for $m\to\infty$.

In order to perform the numerical analysis of the flow equations to 
find the fixed point solution and the related critical exponents
we need to reformulate Eqs. (\ref{eq:uflow}) and (\ref{eq:zflow})
in terms of dimensionless variables. At the same time, to 
simplify the numerical procedure, it is convenient
to get rid of the constant $\alpha$ through the 
transformation already used in the previous Section and,
more important, to replace the differential equation for the potential 
with the corresponding equation for the potential derivative w.r.t. the field.
In fact, as already noticed in \cite{hasen,mor2}
the former equation is stiff and therefore more difficult to treat.
To this aim we introduce the new dimensionless variables 
$t={\rm ln}(\Lambda/k)$, $x=k^{-(1+\eta)/2}\phi/\sqrt{\alpha}$,
$V_d(t,x)=k^{-3} V(k,\phi)/\alpha$ and $Z_d(t,x)=k^{\eta}Z(k,\phi)$
where the subscript $d$ has been used to indicate the dimensionless
potential and wave function renormalization,
$\alpha$ is defined in Eq.(\ref{eq:alfa}) and, 
as before, $\eta$ indicates the anomalous dimension of the field.
Finally we define the potential derivative $f(x,t)=\partial_x V_d(t,x)$.
In terms of these new variables Eqs.(\ref{eq:uflow},\ref{eq:zflow}) 
read
\be
\label{eq:dfzm}
\partial_t f={{(5-\eta)}\over {2}}f - {{(1+\eta)}\over {2}}x f'
-\left ( m-{1\over 2}\right ) \Biggl({{Z_d}\over{Z_d+f'}}\Biggr)^{m-3/2}
{{({Z_d}'f'-Z_d f'')}\over{(Z_d+f')^2}}
\ee
\bea
\label{eq:dzfm}
\partial_t Z_d &=&-\eta Z_d - {{(1+\eta)}\over {2}}x {Z_d}'
+\Biggl({{Z_d}\over{Z_d+f'}}\Biggr)^{m-1/2}
\Biggl\lbrack{{(m-1/2)}\over{Z_d+f'}}
\biggl({Z_d}''-{{49 ({Z_d}')^2}\over{24Z_d}}\biggr)
\nonumber\\
&&-{{7(m^2-1/4)}\over{6 (Z_d+f')^2}}{Z_d}'f''+
{{(m^2-1/4)(m+3/2)}\over{6(Z_d+f')^3}}Z_d(f'')^2\Biggr \rbrack
\eea

Analogously, we can rewrite
Eqs. (\ref{eq:ufminf}) and (\ref{eq:zfminf}) in the dimensionless form
\be
\label{eq:dfz}
\partial_t f ={{(5-\eta)}\over {2}}f - {{(1+\eta)}\over {2}}x f'+ e^{-f'/Z_d}
\left ({{f''}\over{Z_d}}-{{f' {Z_d}'}\over{Z_d^2}}\right )
\ee
\bea
\label{eq:dzf}
\partial_t Z_d=
-\eta Z_d - {{(1+\eta)}\over {2}}x {Z_d}'
+e^{-f'/Z_d}\Biggl({{{Z_d}''}\over{Z_d}}-{{49 ({Z_d}')^2}\over{24Z_d^2}}
-{{7{Z_d}'f''}\over{6 Z_d^2}}+
{{(f'')^2}\over{6 Z_d^2}}\Biggr )
\eea

The fixed points are the $t$-independent solutions $f^*(x)$ and ${Z_d}^*(x)$
of Eqs. (\ref{eq:dfzm},\ref{eq:dzfm}), or Eqs. (\ref{eq:dfz},\ref{eq:dzf}).
The resolution of the coupled ordinary differential 
fixed point equations also provide the value of $\eta$ at the fixed point.
In order to find $\nu$ and $\omega$ we need to introduce small $t$-dependent
perturbations around the fixed point
and reduce the problem  to coupled linear equations in the perturbations.
The perturbations are parametrized in the following way
\bea
\label{eq:pert}
{V_d}(t,x)={V_d}^*(x)+e^{\lambda t}{\rm v}(x)\nonumber \\
Z_d(t,x)={Z_d}^*(x)+e^{\lambda t}z(x)
\eea  
The corresponding relation for the potential derivative $f(t,x)$ is obtained 
by deriving the first line and, 
by indicating with $h(x)$ the derivative of ${\rm v}(x)$, one has
\be
\label{eq:pert2}
f(t,x)=f^*(x)+e^{\lambda t}h(x)
\ee
The resolution of the linear equations provides
the eigenvalues $\lambda$ and the eigenvectors ${\rm v}(x)$, $h(x)$ and $z(x)$
which govern the flow close to the fixed point.
In particular for the problem considered we expect, besides 
the Gaussian fixed point, which corresponds to $f^*(x)=\eta=0,~{Z_d}^*(x)=const$, only
the Wilson-Fisher fixed point, which has just one relevant 
eigenvector ($\lambda$ positive). The exponent $\nu$ is defined
as the inverse of the only positive eigenvalue and $\omega$
is the opposite of the less negative eigenvalue.

We shall also consider the $O(N)$ theory but only in the lowest
order approximation $O(\partial^0)$, and in this case the dimensionless form
of the flow equation for the derivative of the potential is
 \bea
\label{eq:dfmn}
\partial_t f={5\over 2}f - {1 \over 2}x f'
+\left (m-{1\over 2}\right )\Biggl\lbrack\left({1\over{1+f'}}\right)^{m+1/2} f''
\nonumber\\
+(N-1)\left({1\over{ 1+f/x }}\right)^{m+1/2}\left({{f'}\over x}
-{{f}\over {x^2}} \right )\Biggr\rbrack 
\eea
where in addition to the longitudinal field contribution,
which is equal to the one for the single component field, there
are $(N-1)$ contributions of the transverse fields. Finally the limit
$m\to\infty$ of Eq. (\ref{eq:dfmn}) becomes
\be
\label{eq:dfn}
\partial_t f ={5\over 2}f - {1\over 2}x f'+
e^{-f'}f'' +(N-1) e^{-f/x}\left ({{f'}\over x}-{{f}\over {x^2}} \right )
\ee

It must be noted that Eq. (\ref{eq:dfmn}) at $m=1$ and for any $N$ is exactly equal to
the flow equations derived in the $O(\partial^0)$ approximation in \cite{mor2,mor3}
and therefore a good check on our numerical outputs, at least for $m=1$, 
is to compare them to the results reported in  \cite{mor3}.

The numerical resolution of the equations
is thoroughly explained in \cite{mor2}
and here we just shortly summarize the procedure 
and skip the details. 
The two coupled ordinary differential equations  
that determine the fixed point have some boundary 
conditions defined at $x=0$ and the rest at large $x$.
In fact, by requiring the symmetry of the theory under the transformation 
$x\to -x$, one gets the following conditions: ${Z^*_d}'(0)=f^*(0)=0$. In addition, 
the normalization  ${Z^*_d}(0)=1$ is imposed.
(As noticed in \cite{io} the flow equations here considered 
are reparametrization \cite{mor2,comel1} invariant, and, as a consequence 
the universal quantities are independent of the particular normalization
chosen, and this has been checked explicitly in the numerical 
analysis of \cite{io}). 
On the other hand, the equations  themselves constrain the asymptotic 
behavior of the solution at large $x$, up to two constant factors, 
one for each of the two functions  $Z_d^*(x)$ and $f^*(x)$.
The eigenvalue  problem for the linear equations is totally analogous 
because the same symmetry must be imposed at $x=0$ and the same 
kind of constraints can be derived from the equations at large $x$.

These boundary conditions are sufficient to solve 
the problem (and determine the fixed point
and the critical exponents) by means, for instance, 
of the shooting method\cite{mor2}.
In particular, for our analysis, we have implemented this method by 
making use of the NAG libraries and this has allowed us to improve on the 
accuracy of the results of \cite{io} and to test the equations for higher values of $m$.

One result of the numerical analysis has already been presented in Fig. 1,
where the curvature at $x=0$
of the non-Gaussian fixed point solution for some values of $m$,
is plotted. Namely  the triangles in the upper frame of Fig. 1 are the values of $-10~{f^*}'(0)$
determined from Eq. (\ref{eq:dfzm}), in the $O(\partial^0)$ approximation i.e. with
$Z_d=1$ and $\eta=0$, plotted versus $1/m$.
The circles in the upper and lower frames correspond instead to the product
$-10~m {f^*}'(0)$ displayed with  two different scales on the vertical axis.
This plot  has already been commented in Sect. 2.

Fig.2 shows the anomalous dimension $\eta$ at the fixed point for the $N=1$ case,
plotted versus $1/m$ up to $m=120$. Note that we have inserted at $1/m=0$ the value of $\eta$ obtained from the
fixed point solution of Eqs. (\ref{eq:dfz},\ref{eq:dzf})  whereas all the other points come from
Eqs. (\ref{eq:dfzm},\ref{eq:dzfm}). Some of these points are already reported
in \cite{io}, where however the maximum value of $m$ considered was $m=40$.
The convergence in Fig.2 of the values obtained at finite $m$ to the one at $1/m=0$  is clear.

In Fig. 3 we have collected the same kind of results for $\nu$ and $\omega$.
Namely, for the $N=1$ case, empty and black circles are respectively  the $O(\partial^0)$ and  $O(\partial^2)$
values of $\nu$ and  empty and black triangles the same for $\omega$.
Again the points on the vertical axis $1/m=0$ are obtained from Eqs.(\ref{eq:dfz},\ref{eq:dzf})
and the others from  Eqs. (\ref{eq:dfzm},\ref{eq:dzfm}).
Note that for the  $O(\partial^0)$ case we evaluated the exponents up to $m=120$ whereas
for the $O(\partial^2)$ we stopped at $m=10$ because we believe that even at this level
the trend of the data  and their convergence to the $1/m=0$ case is evident.

Two remarks are in order. The first one concerns the fact that,
for those values of $m$ for which the comparison is possible,
we have found agreement with the numerical data of \cite{io},
except for $\omega$ to the  $O(\partial^2)$ order, whose values plotted in Fig. 3 (black triangles)
differ from those of  \cite{io}. In fact, we have noticed that in
this case the results are more sensible to the maximum value of $x$ considered in the numerical
integration of the equations and the routines used here allowed a more careful
analysis and a more accurate determination of the data.
The second remark is about the convergence of the derivative expansion. The
$O(\partial^0)$ and the   $O(\partial^2)$  values of $\nu$ at $1/m=0$ are almost equal,
suggesting that the derivative expansion converges very rapidly for this particular exponent.
The situation for $\omega$ is rather different, and  one can only guess that
the higher order determinations of this quantity lie between the $O(\partial^0)$ and the $O(\partial^2)$
estimates. Therefore it seems that the rapid convergence concerns only the relevant eigenvalue.

The findings of the numerical analysis for $\nu$ and $\omega$ in
the $N=2$ theory and  $O(\partial^0)$  approximation are also reported in
Fig. 3. The symbol $+$ is associated to the values of   $\nu$ and the $\times$ to  $\omega$.
In this case we have determined the exponents up to $m=120$ and again
the convergence toward the values obtained at  $1/m=0$ is clear.
On the basis of these results, we do not consider so many
values of $m$ for  $N=3,4$, and determine $\nu$ and $\omega$
only at  $m=1,2$ and at $1/m=0$ (see the Tables below).

In Fig. 4 we report various plots obtained from
Eq.(\ref{eq:dfn}), which is valid to the $O(\partial^0)$ order,
in the four cases $N=1,2,3,4$. In the upper frame, the
fixed point derivative of the potential is shown, in the central
frame the eigenvector corresponding to the critical exponent $\nu$ and, in the lower
one, the eigenvector corresponding to $\omega$. The trend observed is a flattening
of the curves for increasing $N$.

Finally in Table I and II we report the numerical values of the exponents 
only at $m=1,2,\infty$.
Incidentally we note that $\eta$ at $m=\infty$ in Table I
is practically the same of the value $\eta=0.0329$ quoted in 
\cite{io} as the output of a fit 
to  the data collected  up to $m=40$. We also note that
the $O(\partial^0)$ results at $m=1$ are totally in agreement 
with those of \cite{mor3} which, as discussed above, was expected because in this 
particular case the flow equations considered  here and in \cite{mor3} coincide.

A comparison with the critical
exponents determined by  different methods
(see \cite{zinnj,guida,berg2}) globally indicates 
a better agreement with the values at $m=\infty$
rather than with those at $m=1$ or $m=2$.
This supports the idea that the optimization of the proper time cutoff
is achieved in the large $m$ limit, which can be interpreted as a sharp limit
of this class of regulators. Conversely, the optimization criterion suggested in \cite{litim},
obtained by establishing a relation between the proper time regulator and
the $R_k$ class of regulators introduced in the ERG 
approach, does not seem to be particularly satisfying. 
In fact it predicts that the optimum value of our parameter 
(note the difference in the definition of $m$ here and in \cite{litim})  
is  $m=3/2$ and the exponents at $m=3/2$ are expected to be between those
found at $m=1$ and  those at $m=2$ in Tables I,II.

\section{$O(N)$ theory at $N=\infty$}

Another interesting check on the proper time flow equations concerns
the  problem of the $O(N)$ symmetric theory in the limit 
$N\to \infty$. In this case 
the  $O(N)$ theory reduces to the spherical model which is exactly 
solved. At the same time in this limit
the flow equation for the potential becomes
a close decoupled equation and the main universal features
of the spherical model are recovered
\cite{weg,comel2,mor3}. Our aim is to check that 
these properties survive when we interchange the two limit procedures 
and consider first  $m \to\infty$ and then $N\to\infty$.
As noticed before, to the lowest order in the derivative expansion,
the potential flow equation (\ref{eq:dfmn}) at $m=1$ is equal to the one 
considered in \cite{mor3}, where a detailed analysis of the $N=\infty$
case is presented. So at least for $m=1$ the proper time flow of 
the potential reproduces the results shown in \cite{mor3} and in 
particular the presence, besides the Gaussian fixed point,  
of the Wilson-Fisher fixed point for $D<4$, and the
correct determination of the eigenvalue spectrum of the 
linearized equation in both cases. Then we examine the problem 
at $m\not=1$.

Let us consider the more general version in $D$ dimensions of
Eq.(\ref{eq:dfmn}), which was obtained in $D=3$. For large $N$,
the leading behavior is given by the last term in the r.h.s..
As it has been done for the constant $\alpha$, we can make 
one more rescaling $f\to N f$ and $x\to \sqrt{N}~x$
which allows to take straightforwardly the limit
$N\to \infty$
\be
\label{eq:dtfm}
\partial_t f= \left ({{D+2}\over 2}-{{m}\over{x^2(1+f/x)^{m+1}}}\right )f
- \left ({{D-2}\over 2}-{{m}\over{x^2(1+f/x)^{m+1}}}\right )x~f'
\ee

The fixed point structure is obtained by requiring $\partial_t f=0$.
Actually we are interested in the universal properties at the fixed point
and, as shown in \cite{weg,comel2}, to determine the critical exponents
it is not necessary to have the explicit fixed point solution $f^*(x)$, 
but it is sufficient to determine the essential singularities of the equation.
From the structure of Eq. (\ref{eq:dtfm}) one realizes that there is a non-zero
value of $x$ at which the function $f^*$ vanishes. In fact if at  $x=\overline x$
the second bracket in (\ref{eq:dtfm}) vanishes, then it must be $f^*(\overline x)=0$
since the first bracket at  $x=\overline x$ does not vanish.
Therefore from the second bracket it follows that 
$\overline x=\sqrt{2m/(D-2)}$. We can now expand 
around $\overline x$. By setting in the fixed point equation $x=\overline x+\Delta x$ and 
$f^*(x)= f^{*'}(\overline x)\Delta x+O((\Delta x)^2)$
we get the value of $f^{*'}(\overline x)$:
\be 
\label{eq:fpri}
(m+1) f^{*'}(\overline x) ={4\over{D-2}}-2
\ee
With this information we can analyse the corresponding linearized equation
and, recalling Eq. (\ref{eq:pert2}), we get
\bea
\label{eq:lfm}
&&\lambda h
=\left ({{D+2}\over 2}-{{m}\over{x^2(1+f^*/x)^{m+1}}}\right )h
\nonumber\\
&&-\left ({{D-2}\over 2}-{{m}\over{x^2(1+f^*/x)^{m+1}}}\right )x h'
+{{m(m+1)(f^*-x f^{*'})}\over{x^3(1+f^*/x)^{m+2}}}h
\eea

At $x=\overline x$ the coefficient of $x h'$ in brackets vanishes
and therefore from Eq. (\ref{eq:lfm}) it follows that either 
$h'(\overline x)/h(\overline x)$ 
is singular (with a simple pole singularity) 
or it is finite and in this case the
eigenvalue $\lambda$ is uniquely determined 
(remember that $f^*(\overline x)=0$).
If one requires that the eigenvector $h(x)$
has no singularity and admits the  power expansion  
around $\overline x$,
\be
\label{eq:acca}
h(x)=\sum_{i=n}^{\infty}a_i (x-\overline x)^i
\ee
where the lowest power $n$ is a non-negative integer, 
then one can study the behavior of 
Eq.(\ref{eq:lfm}) close to $x=\overline x$.
The quantity  $h'(x)/h(x)$ is replaced by
the leading singular term $n/(x-\overline x)$
but the pole is canceled by the 
vanishing coefficient of $h'(x)/h(x)$, and therefore
with the help Eq. (\ref{eq:fpri}), at $x=\overline x$
Eq.(\ref{eq:lfm}) is reduced to the simple relation
among the eigenvalue $\lambda$, the non-negative integer $n$
and the number of dimensions $D$
\be
\label{eq:eigen}
\lambda=D-2-2n
\ee
which is the spectrum of the spherical model eigenvalues.

We are also able to derive the eigenvalues relative to the 
Gaussian fixed point. In fact, going back to the linearized equation
(\ref{eq:lfm}), by  evaluating it at the Gaussian fixed point $f^*_G(x)=0$,
and expanding around  $\overline x$, one gets the new spectrum 
\be
\label{eq:eigeng}
\lambda_G=2-(D-2)n
\ee

At the upper critical dimension $D=4$ the two spectra (\ref{eq:eigen}) and 
(\ref{eq:eigeng}) coincide.
Obviously for the simple case of the Gaussian fixed point one could directly
start with the linearized equation of the potential, rather than 
using the potential derivative $f$. By indicating with ${\rm v}_G$ the eigenvector
of the potential, 
the $N=\infty$ flow equation for the potential, linearized around the
Gaussian fixed point $V_G^*=0$, reads
\be
\label{eq:lum}
(\lambda-D){\rm v}_G=\left ({m\over x}- {{(D-2)}\over 2}x\right ){{\rm v}_G}'
\ee
The above equation is solved explicitly
\be
\label{eq:solu}
{\rm v}_G=c\left (x^2 - {{2m}\over{(D-2)}}\right )^{(D-\lambda)/(D-2)}
\ee
where $c$ is the integration constant. If we require 
polynomial (not constant) solutions of the problem, which preserve the 
symmetry $x\to -x$, we have 
$(D-\lambda_G)/(D-2)=n+1$ where $n$ is again a non-negative integer.
Therefore we have found again relation (\ref{eq:eigeng}).
It should be noted that here we have the full solution (\ref{eq:solu})
of Eq. (\ref{eq:lum}) and its parity 
must be explicitly required. Conversely 
this constraint must not be applied to 
the expansion considered above, 
around the  point $\overline x\not=0$.

The relevant point of the above analysis is
that in both  eigenvalue spectra for  the Gaussian 
and the non-trivial fixed point, the dependence on 
the proper time regulator parameter $m$ has disappeared
and the structure of the eigenvalue spectrum is valid 
for any value of $m$. We expect that by performing first the
$m\to \infty$ and then the $N\to \infty$ limit in the flow 
equation, these universal properties are preserved. To prove
this point we perform on Eq. (\ref{eq:dfn}) the same rescaling used to derive
Eq.(\ref{eq:dtfm}) and get
\be
\label{eq:dtf}
\partial_t f= \left ({{D+2}\over 2}\right )f
- \left ({{D-2}\over 2}\right )x~f'+ e^{-f/x}\left ({f'\over{x}}-{f\over{x^2}} \right )
\ee

As before, in the fixed point equation there is a value of the dimensionless field, 
$\overline x=\sqrt{2/(D-2)}$, such that $f^*(\overline x)=0$
and, as can be seen by expanding the fixed point equation
around $\overline x$, ${f^*}'(\overline x)=2(4-D)/(D-2)$.
Then we can analyse  the linearized version of Eq.(\ref{eq:dtf})
around the Gaussian and the non-Gaussian fixed point 
as we did for Eq.(\ref{eq:lfm}) and it is straightforward to derive again
the two equations for $\lambda_G$ and $\lambda$, (\ref{eq:eigeng}) and
(\ref{eq:eigen}), respectively. Therefore, even in this case, 
the flow equation obtained in the limit 
$m\to \infty$ retain the relevant universal properties observed at
$m=1$.

\section{Conclusions}

As already discussed in the Introduction,
there is no proof of convergence of the proper time RG
flow to the full effective action when the infrared scale $k$ is lowered to zero,
so that all modes in the momentum space are effectively taken into account.
On the other hand we have collected here some evidences showing 
the reliability of the predictions of this flow equations
which are certainly comparable to those of the ERG.

The proper time regulator is parametrized by a number $m$, which is 
restricted to integer values. According to \cite{io}, $m$ is related
to the effective width associated to the proper time cutoff and larger values of $m$
correspond to more narrow widths.
We have shown that for a particular redefinition
of the quantities entering the flow equations, it is possible to formally take the limit
$m\to\infty$ and get new differential equations which are totally $m$ independent.

Then, in some numerical examples, we checked that the values of the critical exponents
$\eta$, $\nu$ and $\omega$ as functions of the parameter $m$ are effectively converging
to the numerical values obtained from the asymptotic (in the sense of $m\to\infty$)
equations. As a consequence of this analysis we get an optimized determination
of the above exponents for $m\to\infty$ which are certainly in very good agreement with the
numbers coming from different methods.
Other aspects investigated are the perturbative determination
of the anomalous dimension in $D=4$ and the comparison with the spherical model in
the large $N$ limit. In both cases the asymptotic equations
produce results that totally agree with the ones 
coming from the $m$ dependent flow equations at large $m$.
In the $N=\infty$ case, actually, the universal quantities show no
dependence at all on $m$.

\vfill\eject
\begin{center}
TABLE I
\end{center}
\vspace*{.8 cm}
\begin{table} [h] \centering{
\begin{tabular}{|c||c||c|c||c|c|}  \hline
\multicolumn{6}{|c|}{$N=1$}\\
\hline
&$\eta$&\multicolumn{2}{c||}{$\nu$}&\multicolumn{2}{c|}{$\omega$} \\ 
\hline 
$m$&$O(\partial^2)$ & $O(\partial^0)$ & $O(\partial^2)$ & $O(\partial^0)$ 
&$O(\partial^2)$ \\ 
\hline
$ 1    $&  0.0653 &$  0.6604$&$   0.6337$&$ 0.629$&$  0.677$\\
$ 2    $&  0.0507 &$  0.6439$&$   0.6304$&$ 0.674$&$  0.700$\\  
$\infty$&  0.0330 &$  0.6260$&$   0.6244$&$ 0.762$&$  0.851$\\  
\hline
\end{tabular}
\caption{\em
Critical exponents for $N=1$ at the Wilson-Fisher fixed point 
to the leading, $O(\partial^0)$, and next to leading order, $O(\partial^2)$, 
in the derivative expansion.
}
\label{tab:1}
}
\end{table}
\vspace*{2.5 cm}
\begin{center}
TABLE II
\end{center}
\vspace*{.8 cm}
\begin{table} [h] \centering{
\begin{tabular}{|c||c|c||c|c||c|c||}  \hline
&\multicolumn{2}{c||}{$N=2$}
&\multicolumn{2}{c||}{$N=3$} 
&\multicolumn{2}{c||}{$N=4$}\\ 
\hline 
$m$&$\nu$ & $\omega$ &$\nu$ & $\omega$&$\nu$ & $\omega$\\
\hline
$ 1    $&$ 0.7252 $&$ 0.662$&$  0.7809$&$   0.707$&$ 0.8238$&$  0.751$\\
$ 2    $&$ 0.6990 $&$ 0.681$&$  0.7498$&$   0.700$&$ 0.7924$&$  0.726$\\
$\infty$&$ 0.6689 $&$ 0.746$&$  0.7102$&$   0.734$&$ 0.7481$&$  0.729$\\  
\hline
\end{tabular}
\caption{\em
Critical exponents $\nu$ and $\omega$ for $N=2,3,4$ at the Wilson-Fisher fixed point 
to the leading order $O(\partial^0)$ in the derivative expansion.
}
\label{tab:2}
}
\end{table}

\vfill\eject

\begin{figure}
\psfig{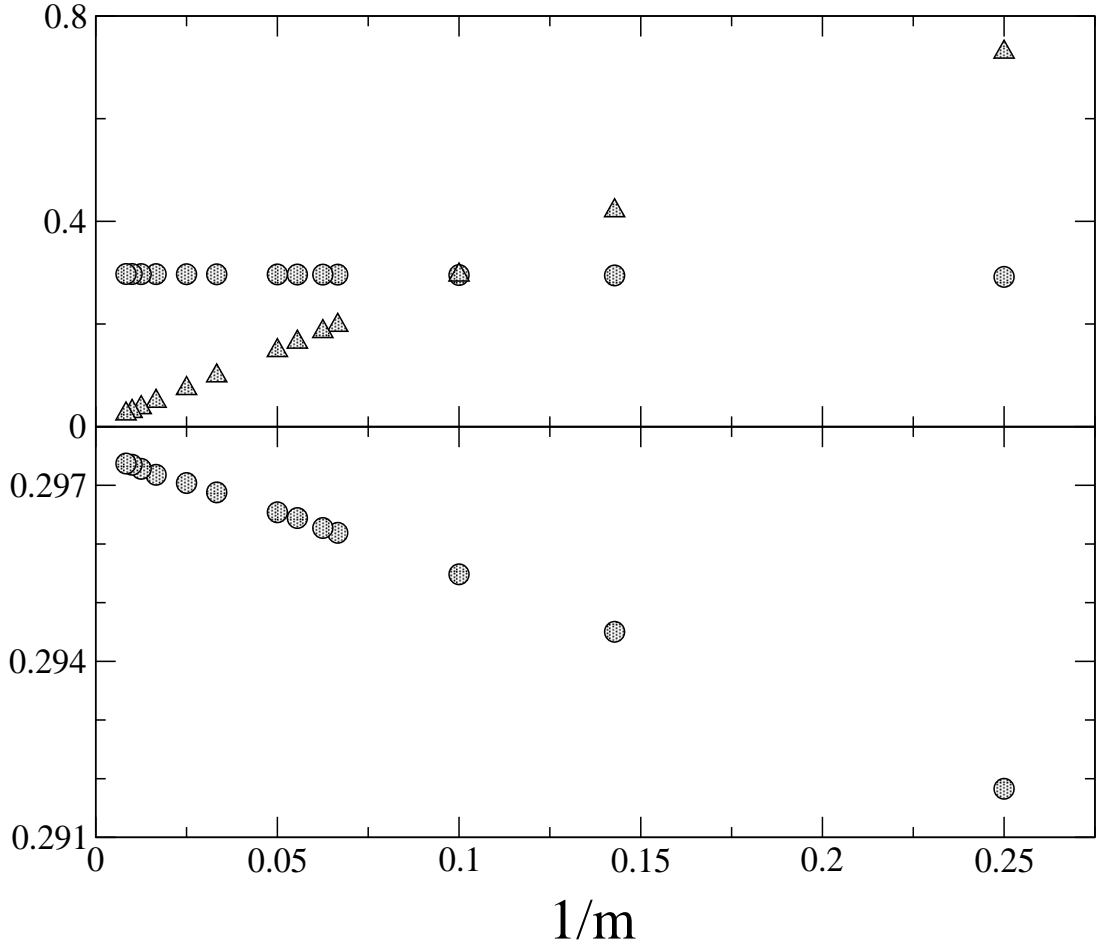}
\caption{
Upper frame: $-10~{f^*}'(0)$ (triangles) and $-10~m {f^*}'(0)$ (circles) at various
values of $1/m$, for $N=1$ and to the leading order in the derivative expansion.
Lower frame: The same circles of the upper frame, plotted with a magnified 
scale in the vertical axis.
}
\end{figure}

\vfill\eject

\begin{figure}
\psfig{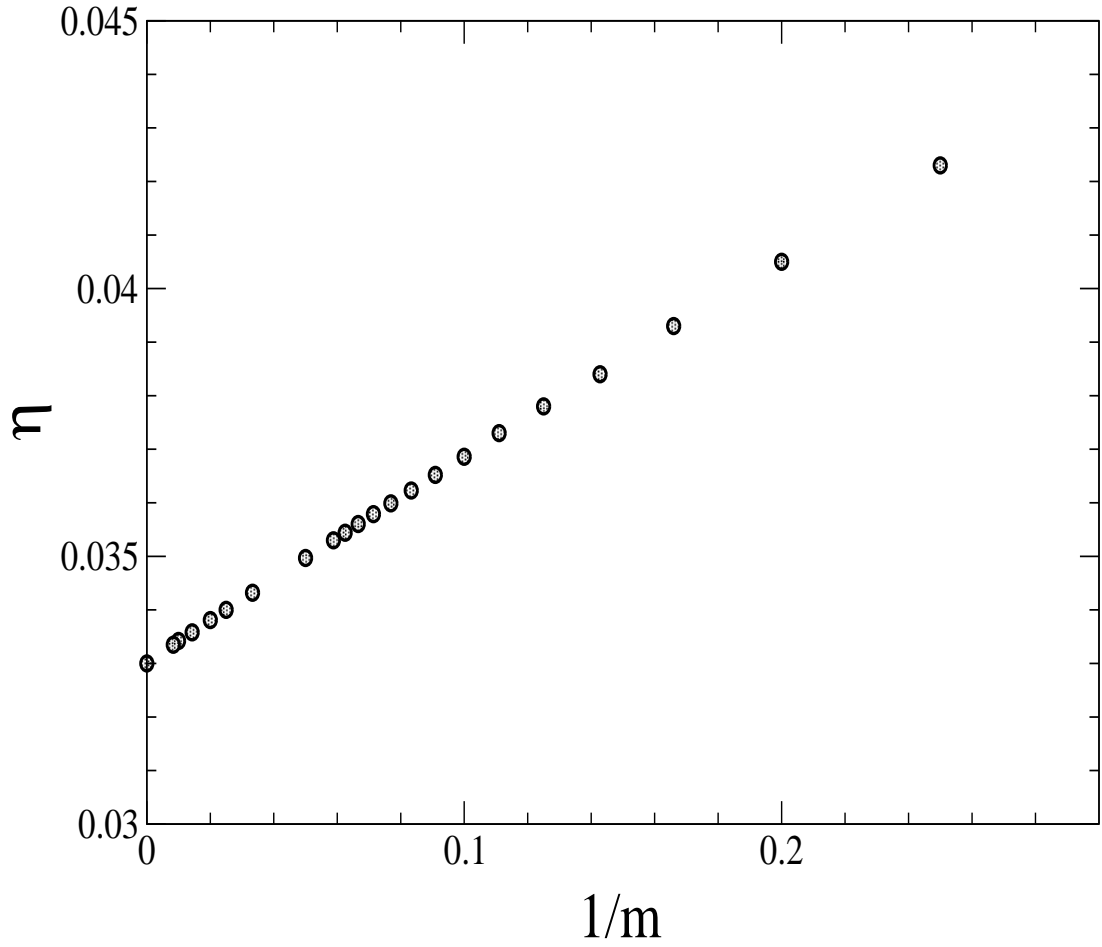}
\caption{
Anomalous dimension $\eta$ at various values of $1/m$, including $1/m=0$,
for $N=1$ and to the next to leading order in the derivative expansion.
}
\end{figure}

\vfill\eject

\begin{figure}
\psfig{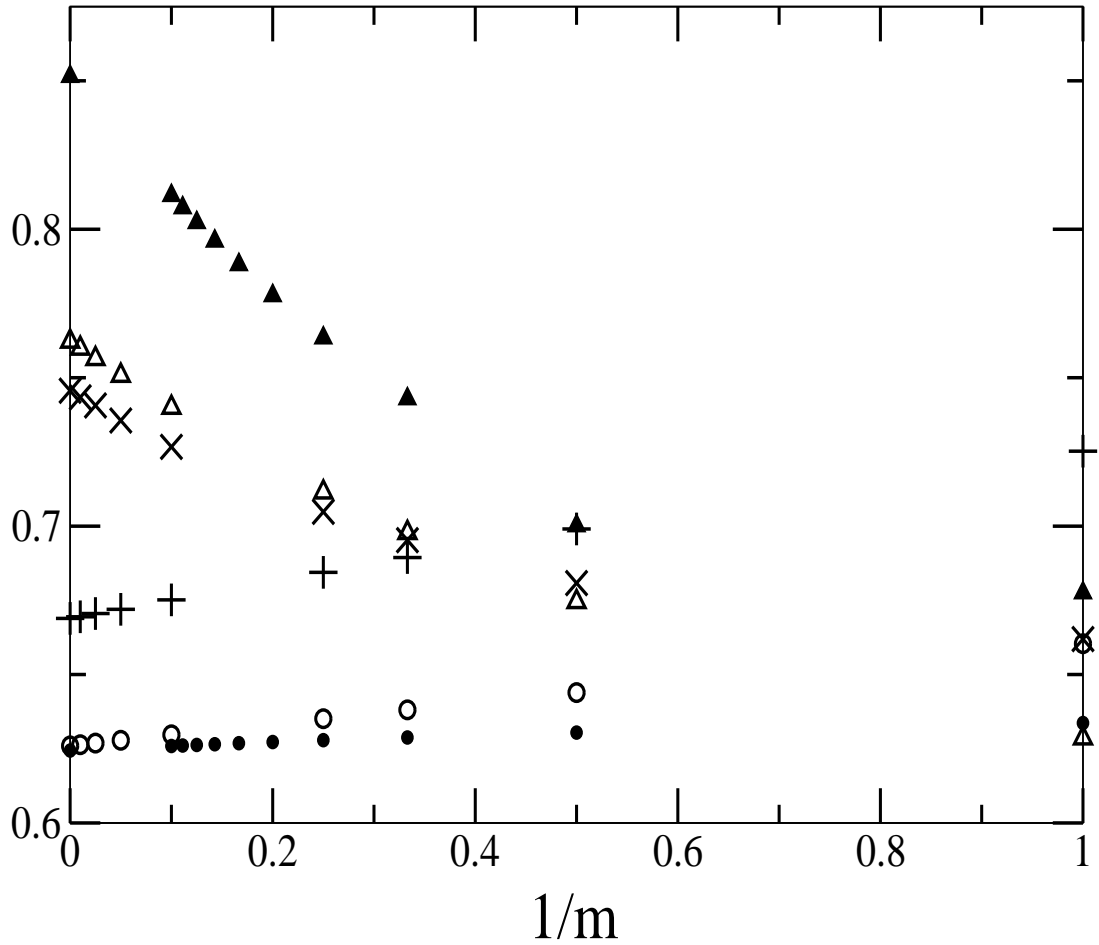}
\caption{
Critical exponents at various values of $1/m$, including $1/m=0$.
Empty and full black circles are respectively
the $O(\partial^0)$  and  $O(\partial^2)$ values of $\nu$ for $N=1$.
Empty and full black triangles  are respectively the $O(\partial^0)$
and  $O(\partial^2)$ values of $\omega$ for $N=1$.
Finally the $+$ and the $\times$ correspond respectively to the $O(\partial^0)$ value of $\nu$
and to the  $O(\partial^0)$ value of $\omega$ for $N=2$.
}
\end{figure}

 \vfill\eject

\begin{figure}
\psfig{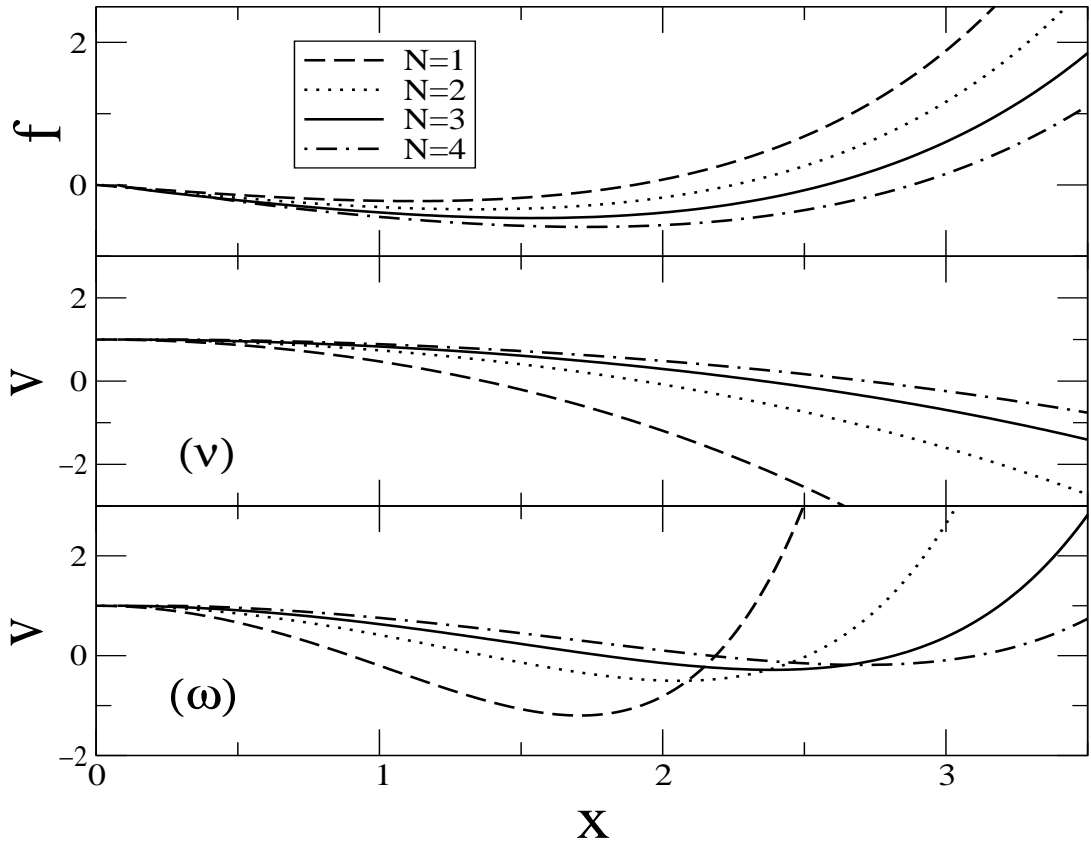}
\caption{
The Wilson-Fisher fixed point derivatives of the potential (upper frame)
and the eigenvectors corresponding to the exponent $\nu$ (central frame)
and those corresponding to the exponent $\omega$ (lower frame), for $N=1,2,3,4$,
all to the  $O(\partial^0)$ order.
}
\end{figure} 
\end{document}